\documentclass[aps,prx,twocolumn,10pt,showpacs,nofootinbib,eqsecnum]{revtex4}
\usepackage{graphicx}
\usepackage{enumerate}
\usepackage{braket}
\usepackage{times,dsfont,amssymb,amsmath,amsthm,amsfonts,amsbsy,mathrsfs,bm,bbm,graphicx,graphics,epsfig,multirow,mathtools,color,bbold}
\usepackage{hyperref}
\hypersetup{
colorlinks=true,
linkcolor=red,
citecolor=red,}

\newcommand{\bdm}[1]{d^{\,2M}\!\bm{#1}\,}

\newcommand{\PQD}{\hbox{PQD}}
\newcommand{\sPQD}{\hbox{($s$)-PQD}}
\newcommand{\SPQD}{\hbox{($\bm{s}$)-PQD}}
\newcommand{\negsPQD}{\hbox{($-s$)-PQD}}
\newcommand{\negSPQD}{\hbox{($-\bm{s}$)-PQD}}
\newcommand{\tPQD}{\hbox{($t$)-PQD}}
\newcommand{\TPQD}{\hbox{($\bm{t}$)-PQD}}

\newcommand{\LON}{\hbox{LON}}

\newcommand{\ssL}{{\scriptscriptstyle L}}
\newcommand{\ssD}{{\scriptscriptstyle D}}
\newcommand{\ssB}{{\scriptscriptstyle B}}
\newcommand{\ssBL}{{\scriptscriptstyle{B\!L}}}
\newcommand{\sszero}{{\scriptstyle 0}}

\newcommand{\Tr}{\text{Tr}}

\definecolor{bw}{rgb}{0.65, 0.16, 0.16}

\begin{document}

\title{Sufficient Conditions for Efficient Classical Simulation of Quantum Optics}

\author{Saleh Rahimi-Keshari}
\affiliation{Centre for Quantum Computation and Communication Technology,
School of Mathematics and Physics, University of Queensland, St Lucia, Queensland 4072, Australia}

\author{Timothy C.~Ralph}
\affiliation{Centre for Quantum Computation and Communication Technology,
School of Mathematics and Physics, University of Queensland, St Lucia, Queensland 4072, Australia}

\author{Carlton M.~Caves}
\affiliation{Center for Quantum Information and Control, University of New Mexico,
MSC07-4220, Albuquerque, New Mexico 87131-0001, USA}
\affiliation{Centre for Engineered Quantum Systems, School of Mathematics and Physics,
University of Queensland, St Lucia, Queensland 4072, Australia}

\date{\today}

\begin{abstract}
We provide general sufficient conditions for the efficient classical simulation of quantum-optics experiments that involve inputting states to a quantum process and making measurements at the output. The first condition is based on the negativity of phase-space quasiprobability distributions (\PQD{s}) of the output state of the process and the output measurements; the second one is based on the negativity of PQDs of the input states, the output measurements, and the transition function associated with the process.  We show that these conditions provide useful practical tools for investigating the effects of imperfections in implementations of boson sampling. In particular, we apply our formalism to boson-sampling experiments that use single-photon or spontaneous-parametric-down-conversion sources and on-off photodetectors.  Considering simple models for loss and noise, we show that above some threshold for the probability of random counts in the photodetectors, these boson-sampling experiments are classically simulatable. We identify mode mismatching as the major source of error contributing to random counts and suggest that this is the chief challenge for implementations of boson sampling of interesting size.

\end{abstract}

\pacs{42.50.Ex, 03.67.Lx, 42.50.-p}


\maketitle

\section{Introduction}

It is generally believed that quantum computers can perform certain tasks faster than their classical  counterparts. Identifying the resource that enables this speedup is of particular interest in quantum information science.  Attempts to identify the elusive quantum feature are generally back-door attacks, studying not what is essential for speedup, but rather what is lacking in quantum circuits that can be efficiently simulated classically.  A promising candidate resource comes from the result that, in general, there is no quantum speedup for circuits whose initial states and operations have nonnegative Wigner functions~\cite{Emer-12,Mari,Emer-13}.  This suggests that negativity of the Wigner function~\cite{Wigner}, which can also be viewed as quantum interference~\cite{Stahlke}, is a necessary resource for quantum speedup.

Of particular interest are quantum-optical models of computation that seem achievable in the near future. There has been considerable interest in boson sampling~\cite{AA} as an intermediate model of quantum computation that, despite its simple physical implementation, is believed to be classically hard.  In this model, $N$ single photons are injected into $N$ ports of an $M$-port ($M\gg N$), lossless, passive linear-optical network (\LON).\footnote{The linear-optical networks considered in the context of boson sampling and within this paper are \emph{passive\/}; i.e., they contain no active elements that generate photons.}  The remaining $M-N$ input ports receive vacuum states.  Using on-off photodetectors at the output of the network, one samples from the output photon-counting probability distribution.  This output distribution, in general, is proportional to the squared modulus of the permanent of a complex matrix.  Computing permanents is a difficult problem, known to be \#P hard in complexity theory~\cite{Valiant,A08}.  In their original proposal, Aaronson and Arkhipov (AA) provided strong evidence that sampling from the output distribution cannot be simulated efficiently classically~\cite{AA}, and this has come to be known as the boson-sampling problem.

Subsequent studies showed that there are other product input states for which sampling from the output probability distribution is classically hard~\cite{RanSam,cat,coh-fock,as-sq}.  This gives rise to questions about the classes of input states and measurements for which sampling the output distribution is classically intractable.  Given the well-developed theory of phase-space quasiprobability distributions (\PQD{s}) for bosonic states and measurements, an inevitable question asks whether negativity of such \PQD{s} is required for classical intractability of the sampling problem.  In addition, a question of both fundamental and practical importance concerns the effect of imperfections on the classical intractability of sampling problems.  There have been various investigations of the effect on boson sampling of imperfections in the \LON~\cite{Lev-Gar,Kal-Kin,Arkhipov,RohRal12} and of mode mismatching~\cite{Shch}.

The present paper makes two contributions to this discussion.  The first, developed in Sec.~\ref{sec:generic}, is to formulate two sufficient conditions for efficient classical simulation of generic quantum-optics experiments: $M$~bosonic modes prepared in an arbitrary bosonic state undergo an $M$-mode (trace-preserving) quantum process; one generates samples by making a measurement on the $M$ output modes (see Fig.~\ref{fig:gen-cir}).  The first condition is based on expressing the probability distribution of the measurement outcomes in terms of a \PQD\ for the output state of the process and \PQD{s} of the elements of the Positive-Operator-Valued Measure (POVM) that describe the output measurement. If the \PQD\ of the output state can be efficiently computed and if for some operator orderings all the \PQD{s} are nonnegative, then efficient classical simulation of sampling is possible. Our second condition generalizes a previous no-go theorem~\cite{Mari}, which was given in terms of the Wigner function, and it is particularly useful when one cannot efficiently compute the \PQD\ of the output state. For this condition, we derive a relation that decomposes the output probability distribution  into three parts: a \PQD\ for the phase-space complex amplitudes of the input state; a transition function associated with the quantum process, which is a conditional \PQD\ for the output complex amplitudes of the process given the input complex amplitudes; and \PQD{s} of the measurement POVM elements. If specific operator orderings exist such that all these \PQD{s}---input, transition, and output---are efficiently describable and nonnegative, sampling from the output probability distribution can be efficiently simulated classically.  These conditions show that negativity is a necessary resource for a generic quantum-optics experiment not to be efficiently simulatable; the result includes boson sampling as the special case where the quantum process is a~\LON.  We emphasize that efficient classical simulation might still be possible using other methods even if our conditions are not satisfied.

Our second contribution, developed in Sec.~\ref{sec:bosonsampling}, is to apply the results of the first one to investigate the effects of imperfections on implementations of boson sampling that use single-photon states~\cite{AA} or two-mode squeezed-vacuum states~\cite{RanSam} as inputs and photodetection at the output.  The imperfections we consider are loss and mode mismatching at the input to and within the \LON\ and subunity efficiency and random counts in the photodetectors.  Considering simple models for these errors, we find necessary and sufficient conditions for the relevant \PQD{s} to be nonnegative, and thus for such boson-sampling implementations to be efficiently simulated classically using these methods.  These conditions say that an experiment can be classically simulated when the probability of random counts per photodetector exceeds some threshold in the experiment.

The various sources of error we consider are not completely independent.  The {\it random counts\/} at the photodetectors include both intrinsic dark counts and counts due to mode-mismatched photons, i.e., nonoverlapping parts of photon wave packets.  These mode-mismatched photons are lost to the interference that gives rise to the desired output photocount distribution. They are part of the losses within the apparatus, but they can make their way through the \LON\ to the photodetectors and be counted within the spatiotemporal windows of the detectors. They thus contribute essentially random counts within the photodetectors.

As we discuss in Secs.~\ref{sec:BSsinglephotons} and~\ref{sec:BSSPDC}, our conditions for classical simulatability are not a challenge for situations with practical losses and high-quality photodetectors, {\it if\/} the only source of random counts is the intrinsic dark counts in the detectors. The chief challenge for boson sampling, we believe, comes when a substantial number of mode-mismatched photons reach the detectors and are counted as random counts.  A good, but not exact rule of thumb is that our methods can classically simulate a boson-sampling experiment when the number of mode-mismatched photons reaching the photodetectors exceeds the number of mode-matched photons.  The analysis in Secs.~\ref{sec:BSsinglephotons} and~\ref{sec:BSSPDC} suggests that this is an important challenge for implementations of boson sampling of interesting size, i.e. when the size of the system is sufficiently large to represent a challenge for a classical computer to sample.

The paper concludes with a discussion of outstanding issues in Sec.~\ref{sec:conclusion}.

\section{Simulation of generic sampling problems}
\label{sec:generic}

\subsection{Generic quantum-optics sampling problem}
\label{sec:genericsampling}

\begin{figure}[tpb]
\centering
\includegraphics[width=0.8\columnwidth]{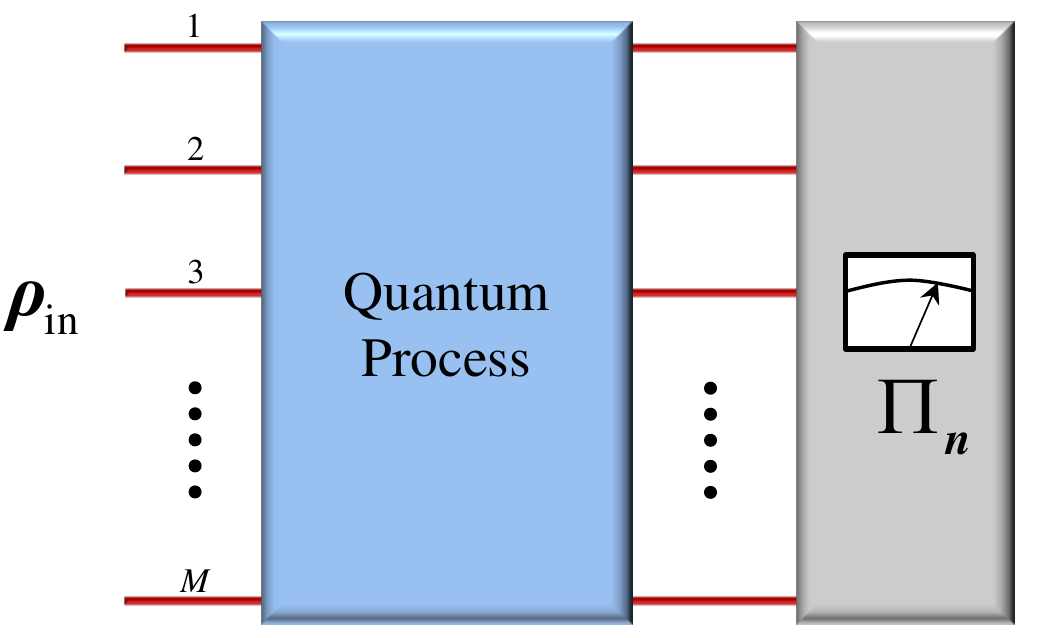}
\caption{Generic quantum-optics sampling problem: The input state $\bm\rho_{\text{in}}$ is processed through an $M$-mode quantum process described by quantum operation $\mathcal{E}$, producing the output state $\bm\rho_{\text{out}}=\mathcal E(\bm\rho_{\text{in}})$; an output probability distribution $p(\bm{n})=\Tr\big[\bm\rho_{\text{out}}\,\Pi_{\bm n}\big]$ is sampled by measuring a POVM $\{\Pi_{\bm n}\}$.}
\label{fig:gen-cir}
\end{figure}

We consider the generic quantum-optics sampling problem depicted in Fig.~\ref{fig:gen-cir}:
$M$ input bosonic modes, with overall density operator $\bm\rho_{\text{in}}$, traverse a quantum process described by a (trace-preserving) quantum operation $\mathcal{E}$, leading to the output state $\bm\rho_{\text{out}}=\mathcal E(\bm\rho_{\text{in}})$; at the output, one measures the $M$ modes, thus sampling from a probability distribution
\begin{align}
\label{eq:gen-pro-dis}
p({\bm n})
=\Tr\big[\bm\rho_{\text{out}}\,\Pi_{\bm{n}}\big],
\end{align}
where POVM elements $\Pi_{\bm{n}}$, with $\bm{n}$ denoting the joint outcome, characterize the measurement.  The POVM satisfies a completeness relation, $\sum_{\bm{n}}\Pi_{\bm{n}}=\mathcal{I}$, with $\mathcal{I}$ denoting the identity operator in the $M$-mode Fock space.  The quantum operation $\mathcal{E}$ is a linear, trace-preserving, completely positive map from the set of all density operators associated with a quantum system to itself. In general, such a quantum operation describes the system dynamics associated with a joint unitary transformation on the system and an environment and arises formally from tracing out the environment~\cite{NilChu}.

The question is whether sampling from the output probability distribution (\ref{eq:gen-pro-dis}) can be efficiently simulated on a classical computer.  If such classical simulation is possible, then using Stockmeyer's approximate counting algorithm~\cite{Stockmeyer}, one can approximate the output probability to within a multiplicative error in $\text{BPP}^{\text{NP}}$, which is contained in the third level of the polynomial hierarchy; $\tilde{p}(\bm{n})$ approximates $p(\bm{n})$ to within a multiplicative factor $g$ if $p(\bm{n})/g\leq\tilde{p}(\bm{n})\leq p(\bm{n}) g$.

Ideal boson sampling is a special case of this general problem, for which the input state is a multimode Fock state with $N$ single photons, the quantum process is a lossless LON described by an $M\times M$ unitary matrix $\bm{U}$ with $M\gg N$, and photon-counting measurements are made on each output mode.  The output probabilities, in general, are proportional to the squared modulus of permanents of complex matrices, which are, in the likely event of single-photon detections, submatrices of $\bm{U}$~\cite{Scheel}.  Computing permanents is a difficult problem, known to be \#P hard~\cite{Valiant,A08}, and in a class that contains the entire polynomial hierarchy of complexity classes~\cite{Toda}.

The key observation by Aaronson and Arkhipov (AA) was that multiplicative approximation of the squared modulus of permanents of real matrices is also a \#P-hard problem, and it is likely this is the case for general complex matrices~\cite{AA}.  If boson sampling were classically simulatable, one could use Stockmeyer's approximate counting algorithm to approximate the output probability to within a multiplicative error, and this would lead to the collapse of the polynomial hierarchy to the third level, which is thought to be unlikely~\cite{AA}.  Given two plausible conjectures, AA further showed that it is likely that classical simulation of sampling from probability distributions that closely approximate the ideal output probability, known as {\it approximate\/} boson sampling, is also hard.

The approximate sampling problem is of more practical interest than {\it exact\/} sampling because in an experiment the input quantum state, quantum process, and output measurement are only characterized approximately, so one does not sample from the exact probability distribution (\ref{eq:gen-pro-dis}). Moreover, one might not be able to distinguish efficiently sampling from two probability distributions that are close to one another.  Hence, in practice, an interesting sampling problem is the one for which approximate sampling is hard.  In this paper, we do not consider this form of sampling; instead we focus on simulating exact sampling from output distributions that arise in the presence of errors and imperfections. Even though we do not consider approximate sampling explicitly, our simulation methods for exact sampling do lead to a sufficient condition for approximate sampling to be classically simulatable.

We motivate our methods for classical simulation by considering a simple special case of the generic sampling problem.  Suppose the multimode input state $\bm\rho_{\text{in}}$ has a nonnegative Glauber-Sudarshan $P$ function~\cite{Glauber,Sudarshan} $P({\bm\alpha}|\bm\rho_{\text{in}})$, i.e.,
\begin{align}
\bm\rho_{\text{in}}=\int\bdm{\alpha}P({\bm\alpha}|\bm\rho_{\text{in}})\ket{\bm\alpha}\!\bra{\bm\alpha};
\end{align}
here $\ket{\bm\alpha}$ is an $M$-mode coherent state with phase-space complex amplitudes $\bm\alpha=(\alpha_1,\alpha_2,\ldots,\alpha_M)$.\footnote{Throughout vectors are row vectors.  For vectors of complex numbers, e.g., $\bm\alpha,\bm\beta,\bm\xi,\bm\zeta$, the dagger transposes to a column vector and takes a complex conjugate; for the vector of annihilation operators, $\bm{a}$, the dagger transposes and takes the adjoint.}  Such states, as mixtures of coherent states, are often called classical states.  Suppose further that the quantum process transforms multimode coherent states to classical states; such processes are known as classical processes~\cite{ProNonCla}.  Then the output state $\bm\rho_{\text{out}}$ is classical as well and has a nonnegative $P$ function.  Using the linearity of quantum processes over density operators, the output state can be expressed as
\begin{align}
\begin{split}
\bm\rho_{\text{out}}
&=\int\bdm{\alpha}\,\mathcal{E}\big(\ket{\bm\alpha}\!\bra{\bm\alpha}\big)
P(\bm\alpha|\bm\rho_{\text{in}})\\
&=\int\bdm{\beta}\ket{\bm\beta}\!\bra{\bm\beta}
\int\bdm{\alpha}\,P_{\mathcal{E}}(\bm\beta|\bm\alpha)\,P(\bm\alpha|\bm\rho_{\text{in}}),
\end{split}
\end{align}
where $P_{\mathcal{E}}(\bm\beta|\bm\alpha)$ is the $P$ function of the state $\mathcal{E}\big(\ket{\bm\alpha}\!\bra{\bm\alpha}\big)$.  Hence, the output probability distribution~(\ref{eq:gen-pro-dis}) is given by
\begin{align}
\label{eq:pro-cl-cl}
p(\bm{n})=\int\!\bdm{\beta}\,\pi^{M}Q_{\Pi}(\bm{n}|\bm\beta)
\int\!\bdm{\alpha} P_{\mathcal{E}}(\bm\beta|\bm\alpha)\,P(\bm\alpha|\bm\rho_{\text{in}}),
\end{align}
where the Husimi $Q$ functions~\cite{Husimi} of the POVM elements, $Q_{\Pi}(\bm{n}|\bm\beta)=\bra{\bm\beta}\Pi_{\bm{n}}\ket{\bm\beta}/\pi^M$, are always nonnegative and satisfy $\pi^M \sum_{\bm{n}} Q_{\Pi}(\bm{n}|\bm\beta)=1$.

As all the \PQD{s} in the expression~(\ref{eq:pro-cl-cl}) are nonnegative, sampling from the output photon-counting probability distribution can be efficiently simulated on a classical computer, provided the \PQD{s} can be efficiently generated: $\bm\alpha$ is chosen from the input probability distribution $P(\bm\alpha|\bm\rho_{\text{in}})$; given $\bm\alpha$, $\bm\beta$ is chosen according to the probability distribution $P_{\mathcal{E}}(\bm\beta|\bm\alpha)$; finally, given $\bm\beta$, $\bm{n}$ is chosen according to the measurement distribution $\pi^M Q_{\Pi}(\bm{n}|\bm\beta)$.  Applying this procedure to input thermal states for \LON{s} and using Stockmeyer's approximate counting algorithm, it was shown that permanents of Hermitian positive-semidefinite matrices can be approximated to within a multiplicative constant in~$\text{BPP}^{\text{NP}}$~\cite{thermal}.

In the next subsection, we generalize this approach to nonclassical states and processes.  The key idea, taken over from classical states and processes, is to use phase-space complex amplitudes and associated \PQD{s} to describe the input state, the transition through the quantum process, and the output measurements.  The generalization is to use operator orderings more general than the normal ordering of the $P$ function and the antinormal ordering of the $Q$ function, thus expanding the possibilities for finding nonnegative \PQD{s}.

\subsection{Sufficient conditions for efficient classical simulation}
\label{sec:sufficient}

To formulate our condition for efficient classical simulation of the generic sampling problem, we use the
$\bm s$-ordered phase-space quasiprobability distributions [\SPQD{s}] of a Hermitian operator $\bm\rho$, which are defined by~\cite{CahGla69,HilOCo84}
\begin{align}
\label{eq:s-PQD-def}
W^{(\bm s)}(\bm\beta|\bm\rho)
=\int\frac{\bdm{{\xi}}}{\pi^{2M}}\,
\Phi^{(\bm s)}(\bm\xi|\bm\rho)\,
e^{\bm\beta\bm\xi^{\dagger}-\bm\xi\bm\beta^{\dagger}},
\end{align}
where
\begin{align}
\label{eq:s-cha-def}
\Phi^{(\bm{s})}(\bm\xi|\bm{\rho})=\Tr\big[\bm\rho D(\bm\xi)\big]e^{\bm\xi\bm{s}\bm\xi^\dagger/2}
\end{align}
is the corresponding $\bm{s}$-ordered characteristic function, with
\begin{align}
D(\bm\xi)=e^{\bm\xi\bm a^\dagger-\bm a\bm\xi^\dagger}=
\bigotimes_{j=1}^M D(\xi_j)
\end{align}
being the $M$-mode displacement operator, $\bm a=( a_1,\ldots,a_M)$ the row vector of modal annihilation operators, and $\bm a^\dagger=(a_1^\dagger,\ldots,a_M^\dagger)^T$ the column vector of creation operators.  The diagonal matrix $\bm s=\text{diag}(s_1,s_2,\ldots,s_M)$ has the various ordering parameters on the diagonal.

Equation~(\ref{eq:s-PQD-def}) is a Fourier transform, which can be inverted using
\begin{align}\label{eq:Ddelta}
\int\frac{\bdm{{\beta}}}{\pi^{2M}}\,
e^{\bm\zeta\bm\beta^{\dagger}-\bm\beta\bm\zeta^{\dagger}}
=\delta^{2M}(\bm{\zeta})
\end{align}
to give
\begin{align}
\label{eq:s-cha-def2}
\Phi^{(\bm s)}(\bm\xi|\bm\rho)
=\int\bdm{{\beta}}
W^{(\bm s)}(\bm\beta|\bm\rho)\,
e^{\bm\xi\bm\beta^{\dagger}-\bm\beta\bm\xi^{\dagger}}.
\end{align}

Because $\bm\rho$ is Hermitian, the characteristic function satisfies $\Phi^{(\bm{s})*}(\bm\xi|\bm{\rho})=\Phi^{(\bm{s})}(-\bm\xi|\bm{\rho})$, and the $\SPQD$~(\ref{eq:s-PQD-def}) is real.  The $\SPQD$ $W^{(\bm s)}(\bm\beta|\bm\rho)$ gives the $M$-mode Husimi $Q$ function, the Wigner function, and the Glauber-Sudarshan $P$ function for $\bm s=-\bm{I}_M$, $\bm s=0$, and $\bm s=\bm{I}_M$, respectively, where $\bm{I}_M$ denotes the $M\times M$ identity matrix. It is easy to check that the $\SPQD$s are normalized according to
\begin{align}
\int\bdm{{\beta}} W^{(\bm s)}(\bm\beta|\bm\rho)=\Tr[\bm\rho].
\end{align}
These are usually called \PQD{s} when $\bm\rho$ is a density operator, but we generalize the terminology to any Hermitian operator so we can apply it to POVM elements.

The outcome probabilities~(\ref{eq:gen-pro-dis}) can be expressed in terms of the \PQD{s} of the output state and the POVM as~\cite{CahGla69}
\begin{align}
\label{eq:Pn-PQD}
p(\bm{n})
=\int\bdm{\beta}\,\pi^M W^{(-\bm s)}_{\Pi}(\bm{n}|\bm\beta)\,W^{(\bm s)}(\bm\beta|\bm\rho_{\rm out}),
\end{align}
where the measurement $\negSPQD$ is
\begin{align}
\label{eq:s-Pi-PQD-def}
W_\Pi^{(-\bm s)}(\bm n|\bm\beta)
=\int\frac{\bdm{{\xi}}}{\pi^{2M}}\,
\Tr\big[\Pi_{\bm n}D(\bm\xi)\big]\,
e^{-\bm\xi\bm{s}\bm\xi^\dagger/2}\,
e^{\bm\beta\bm\xi^{\dagger}-\bm\xi\bm\beta^{\dagger}}.
\end{align}
These measurement $\negSPQD$s are normalized according to
\begin{align}
\pi^{M}\sum_{\bm{n}}W^{(-\bm s)}_{\Pi}(\bm{n}|\bm\beta)=1,
\end{align}
for any values of $\bm\beta$ and $\bm s$,  as one sees by applying $\Tr[D(\bm\xi)]=\pi^M\delta^{2M}(\bm\xi)$ to Eq.~(\ref{eq:s-Pi-PQD-def}).

{\it First condition.} We now present a sufficient condition for efficient classical simulation of the sampling problem.  If there exist values of $\bm s$ such that the \PQD{s} in Eq.~(\ref{eq:Pn-PQD}) are nonnegative, they can be used to simulate sampling from $p(\bm{n})$ in two steps:
\begin{enumerate}[(i)]
\item The vector of complex amplitudes $\bm\beta$ is chosen from the probability distribution $W^{(\bm s)}(\bm\beta|\bm\rho_{\rm out})$.
\item For the given $\bm\beta$, the outcome $\bm{n}$ is chosen from the probability distribution $\pi^{M} W^{(-\bm s)}_{\Pi}(\bm{n}|\bm\beta)$.
\end{enumerate}
This condition is particularly useful if the \SPQD\ of the output state can be efficiently computed, as it can be, for example, for Gaussian input states and Gaussian processes~\cite{Gaussian}.

For cases where efficient computation of the output \SPQD\ is not possible, we now derive a general expression that relates the \SPQD\ of the output state $\bm\rho_\text{out}$ to the \TPQD\ of the input state $\bm\rho_\text{in}$.  This derivation introduces the transfer function, which transfers complex amplitudes from input to output of the quantum process and which depends on both the input and output operator \hbox{orderings}.

An $M$-mode input state can be expanded in terms of displacement operators,
\begin{align}
\bm\rho_{\text{in}}=\int\frac{\bdm{{\xi}}}{\pi^M}\, \Phi^{(\bm{t})}(\bm\xi|\bm\rho_{\text{in}})\, e^{-\bm\xi\bm{t}\bm\xi^{\dagger}/2} D^\dagger(\bm\xi).
\end{align}
In this expression, we can replace $D^\dagger(\bm\xi)=D(-\bm\xi)$ with $D(\bm\xi)$ if we wish, because $\bm\rho_{\text{in}}$ is Hermitian.  Linearity of quantum processes implies that
\begin{align}
\bm\rho_{\text{out}}=
\int \frac{\bdm{{\xi}}}{\pi^M}\,\Phi^{(\bm{t})}(\bm\xi|\bm\rho_{\text{in}})\, e^{-\bm\xi\bm{t}\bm\xi^{\dagger}/2}\,\mathcal{E}\big(D^\dagger(\bm\xi)\big),
\end{align}
from which we obtain the $(\bm s)$-ordered characteristic function of the output state,
\begin{widetext}
\begin{align}
\Phi^{(\bm{s})}(\bm\zeta|\bm\rho_{\text{out}})
=\int\frac{\bdm{{\xi}}}{\pi^M}\,
\Phi^{(\bm{t})}(\bm\xi|\bm\rho_{\text{in}})\, e^{-\bm\xi\bm{t}\bm\xi^{\dagger}/2+\bm\zeta\bm{s}\bm\zeta^{\dagger}/2}
\,\Tr\big[\mathcal{E}\big(D^\dagger(\bm\xi)\big)D(\bm\zeta)\big].
\end{align}
Using the Fourier transform~(\ref{eq:s-PQD-def}) and its inverse~(\ref{eq:s-cha-def2}), we can obtain the relation between the input and output \PQD{s},
\begin{align}
\label{eq:out-in-PQDs}
W^{(\bm s)}(\bm\beta|\bm\rho_\text{out})
=\int \bdm{{\alpha}}
T_{\mathcal{E}}^{(\bm s,\bm t)}(\bm\beta|\bm\alpha)\,W^{(\bm t)}(\bm\alpha|\bm\rho_\text{in}),
\end{align}
where the transition function associated with the quantum process is defined by
\begin{align}\label{trans-fun}
T_{\mathcal{E}}^{(\bm s,\bm t)}(\bm\beta|\bm\alpha)
=\int\frac{\bdm{{\zeta}}}{\pi^{2M}}\,
e^{\bm\zeta\bm{s}\bm\zeta^{\dagger}/2}\,e^{\bm\beta\bm\zeta^\dagger-\bm\zeta\bm\beta^\dagger}
\!\int\frac{\bdm{{\xi}}}{\pi^M}\,
e^{-\bm\xi\bm{t}\bm\xi^{\dagger}/2}\,
e^{\bm\xi\bm\alpha^\dagger-\bm\alpha\bm\xi^\dagger}\,
\Tr\big[\mathcal{E}\big(D^\dagger(\bm\xi)\big)D(\bm\zeta)\big].
\end{align}

The quantity $\Tr\big[\mathcal{E}\big(D^\dagger(\bm\xi)\big)D(\bm\zeta)\big]$ gives the ``matrix elements'' of the quantum process $\mathcal E$ in the displacement-operator basis.  We can use the antinormally ordered form of the displacement operator, combined with the coherent-state resolution of the identity, $\mathcal{I}=\int\bdm{{\gamma}}\ket{\bm\gamma}\!\bra{\bm\gamma}/\pi^M$, to obtain
\begin{align}
\label{eq:Dantinormal}
e^{-\bm\xi\bm\xi^\dagger/2}D(\bm\xi)
=e^{-\hat{\bm{a}}\bm\xi^\dagger}\mathcal{I}e^{\bm\xi\hat{\bm{a}}^{\dagger}}
=\int\frac{\bdm{{\gamma}}}{\pi^M}\,
\ket{\bm\gamma}\!\bra{\bm\gamma}\,
e^{\bm\xi\bm\gamma^{\dagger}-\bm\gamma\bm\xi^\dagger}.
\end{align}
This allows us to convert $\mathcal{E}\big(D^\dagger(\bm\xi)\big)$ into the action of the quantum process on coherent states:
\begin{align}
\label{eq:pros-on-coh}
\mathcal{E}\big(D^\dagger(\bm\xi)\big)
=e^{\bm\xi\bm\xi^\dagger/2}
\int\frac{\bdm{\gamma}}{\pi^M}\,
e^{\bm\gamma\bm\xi^\dagger-\bm\xi\bm\gamma^{\dagger}}
\mathcal{E}\big(\ket{\bm\gamma}\!\bra{\bm\gamma}\big).
\end{align}
Using Eqs.~(\ref{trans-fun}) and~(\ref{eq:pros-on-coh}), one can check that for trace-preserving quantum processes, we have
\begin{align}
\int \bdm{\beta} T_{\mathcal{E}}^{(\bm s,\bm t)}(\bm\beta|\bm\alpha)=1.
\end{align}
We do not plug Eq.~(\ref{eq:pros-on-coh}) into Eq.~(\ref{trans-fun}) because generally the integrals diverge; the art of finding a well-behaved transition function is, for a specific quantum process, to find the most favorable input and output ordering parameters, $\bm s$ and $\bm t$, that make the integrals converge.

Now, by combining Eqs.~(\ref{eq:Pn-PQD}) and~(\ref{eq:out-in-PQDs}), we can assemble the ingredients for a classical simulation of sampling from the output distribution of the quantum circuit shown in Fig.~\ref{fig:gen-cir},
\begin{align}
\label{eq:Pn-PQD-tran}
p(\bm{n})
=\int\bdm{\beta}\!\!\int\bdm{{\alpha}}
\pi^M W^{(-\bm s)}_{\Pi}(\bm{n}|\bm\beta)\,
T_\mathcal{E}^{(\bm s,\bm t)}(\bm\beta|\bm\alpha)\,
W^{(\bm t)}(\bm\alpha|\bm\rho_\text{in}).
\end{align}
\end{widetext}
\noindent

{\it Second condition.} We can carry out a classical simulation, using the following procedure, if there exist values of $\bm t$ and $\bm s$ such that the \PQD{s} of the input, the transition function, and the measurement are all nonnegative and efficiently describable:
\begin{enumerate}[(i)]
\item The vector of complex amplitudes $\bm\alpha$ is chosen from the input probability distribution $W^{(\bm t)}(\bm\alpha|\bm\rho_\text{in})$.
\item For the given $\bm\alpha$, the vector $\bm{\beta}$ is chosen from the transition probability distribution $T_\mathcal{E}^{(\bm s,\bm t)}(\bm\beta|\bm\alpha)$.
\item For the given $\bm\beta$, the outcome $\bm{n}$ is chosen from the output probability distribution $\pi^{M} W^{(-\bm s)}_{\Pi}(\bm{n}|\bm\beta)$.
\end{enumerate}
\noindent
That the three probability distributions are efficiently describable must be judged on a case-by-case basis.  For the input, this is generally achieved by assuming that the input state $\bm\rho_\text{in}$ is a product state of the $M$ modes or, perhaps, a product of blocks of modes of fixed size; likewise, for the output, the measurements are generally product measurements of the $M$ modes or products of measurements on blocks of fixed size.  The complexity of the transition function depends on the quantum process; for the \LON{s} used in boson sampling, the transition function is Gaussian and can be \hbox{generated} \hbox{trivially} from the matrix that describes the \LON, as we show in Sec.~\ref{sec:bosonsampling}.

This second condition includes the previous results as special cases.  For classical states and classical processes, we can choose $\bm s=\bm t=\bm{I}_M$, and Eq.~(\ref{eq:Pn-PQD-tran}) reduces to Eq.~(\ref{eq:pro-cl-cl}).  In addition, for $\bm s=\bm t=0$, we have that when the transition function is nonnegative and the input quantum state and output POVM elements have nonnegative Wigner functions, sampling from the output distribution can be simulated classically~\cite{Mari}.

A procedure for determining if there are input and output orderings that give nonnegative quasidistributions is the following.  The \TPQD\ $W^{(\bm t)}(\bm\alpha|\bm\rho_\text{in})$ of the input state is nonnegative for ordering parameters $\bm t\le\bar{\bm t}$, where $\bar{\bm t}\ge-{\bm I}_M$, and the \negSPQD\ $W_\Pi^{(-\bm s)}(\bm n|\bm\beta)$ of the POVM elements is nonnegative for $\bm s\ge\bar{\bm s}$, i.e., $s_k\ge \bar{s}_k$, for all $k$, where $\bar{\bm s}\le{\bm I}_M$.  The necessary and sufficient condition for there also to be a nonnegative transition function is that $T^{(\bar{\bm s},\bar{\bm t})}(\bm\beta|\bm\alpha)$ be nonnegative and no more singular than a $\delta$ function.  This is a necessary and sufficient condition for our second condition to yield a classical simulation.

A crucial point, in general and for what follows, is that simulations using our first condition provide tighter bounds for classical simulation than the second condition because the first condition, unlike the second, does not require anything about the input \PQD, in particular, that it be nonnegative. To make the difference clear, suppose the input state is a nonclassical Gaussian state, and the process is also a nonclassical process, but one that cancels the nonclassicality of the input state in such a way as to make the output state classical; an example is provided by input squeezed states and antisqueezing operations.  In this situation, for any measurement at the output, the experiment is classically simulatable according to the first condition.  For the second condition, however, the input \PQD\ and the transition function are only nonnegative for a limited range of ordering parameters $\bm t\le\bar{\bm t}< \bm{I}_M$, and only for certain measurements does the second condition hold.

\vspace{12pt}
\section{Efficient classical simulations of implementations of boson sampling}
\label{sec:bosonsampling}

\subsection{General considerations for passive linear optical networks}
\label{sec:pLON}

We now consider the case where the quantum process is a lossy $M$-mode \LON. In this case the quantum process takes coherent states to coherent states according to~\cite{charact}
\begin{align}
\label{eq:E-LON}
\mathcal{E}_{\text{\LON}}\big(\ket{\bm\gamma}\!\bra{\bm\gamma}\big)
=\ket{\bm\gamma\bm{L}}\!\bra{\bm\gamma\bm{L}},
\end{align}
where $\bm{L}$ is the $M\times M$ transfer matrix describing the \LON.  A~\LON\ is an example of what we call a classical process in Sec.~\ref{sec:genericsampling}.

For a lossless \LON, the matrix $\bm L$ is the unitary matrix $\bm U$ mentioned previously.  When there are losses, the quantum operation~(\ref{eq:E-LON}) follows from a very simple model of an environment: In addition to the $M$ actual modes, there are $M$ loss (environment) modes that are initially in vacuum and that carry away photons lost within the \LON; the larger \LON\ that includes the loss modes is described by a unitary operator $\tilde{\mathcal{U}}$, which transforms annihilation operators according to
\begin{align}
\tilde{\mathcal{U}}^\dagger
\begin{pmatrix}
\bm a&\bm a_0
\end{pmatrix}
\tilde{\mathcal{U}}
=\begin{pmatrix}
\bm a&\bm a_0
\end{pmatrix}\bm\tilde{\bm U},
\end{align}
where $\bm a_0$ is the row vector of annihilation operators for the $M$ loss modes and
\begin{align}
\tilde{\bm U}=
\begin{pmatrix}
\bm L&\bm N\\\bm P&\bm M
\end{pmatrix}
\end{align}
is the unitary matrix that describes the complex-amplitude transformation within the larger \LON.  The larger \LON\ takes overall coherent states to overall coherent states according to
\begin{align}
\tilde{\mathcal{U}}
\big|
\begin{pmatrix}
\bm\gamma&\bm\gamma_0
\end{pmatrix}
\big\rangle
=\big|
\begin{pmatrix}
\bm\gamma&\bm\gamma_0
\end{pmatrix}
\tilde{\bm U}
\big\rangle.
\end{align}
The quantum operation~(\ref{eq:E-LON}) follows from tracing out the loss modes:
\begin{align}
\begin{split}
\mathcal{E}_{\text{\LON}}\big(\ket{\bm\gamma}\!\bra{\bm\gamma}\big)
&=\Tr_0\bigl[
\tilde{\mathcal{U}}
\bigl|
\begin{pmatrix}
\bm\gamma&\bm0
\end{pmatrix}
\bigr\rangle
\bigl\langle
\begin{pmatrix}
\bm\gamma&\bm0
\end{pmatrix}
\bigr|
\tilde{\mathcal{U}}^\dagger
\bigr]\\
&=\Tr_0\bigl[
\bigl|
\begin{pmatrix}
\bm\gamma\bm L&\bm\gamma\bm N
\end{pmatrix}
\bigr\rangle
\bigl\langle
\begin{pmatrix}
\bm\gamma\bm L&\bm\gamma\bm N
\end{pmatrix}
\bigr|
\bigr]\\
&=\ket{\bm\gamma\bm{L}}\!\bra{\bm\gamma\bm{L}}.
\end{split}
\end{align}
What the model teaches is that $\bm L$ is a submatrix of the larger unitary matrix $\tilde{\bm U}$ and thus satisfies $\bm L^\dagger\bm L=\bm I_M-\bm P^\dagger\bm P\le\bm I_M$.  In an experiment, the transfer matrix $\bm L$ of any \LON\ can be efficiently characterized by inputting coherent states~\cite{charact}.

\begin{widetext}
We can use the normally ordered form of the displacement operator, $D(\bm{\zeta})=e^{-\bm\zeta\bm\zeta^\dagger/2}e^{\bm\zeta\bm a^\dagger}e^{-\bm a\bm\zeta^\dagger}$, to obtain
\begin{align}
\Tr\big[\mathcal{E}_{\text{\LON}}\big(\ket{\bm\gamma}\!\bra{\bm\gamma}\big)D(\bm\zeta)\big]
=\Tr\big[\ket{\bm\gamma\bm L}\!\bra{\bm\gamma\bm{L}} D(\bm\zeta)\big]=e^{-\bm\zeta\bm\zeta^\dagger/2} e^{\bm\zeta\bm{L}^\dagger\bm\gamma^\dagger-\bm\gamma\bm{L}\bm\zeta^\dagger}.
\end{align}
Plugging this into Eq.~(\ref{eq:pros-on-coh}) and invoking Eq.~(\ref{eq:Ddelta}) gives us
\begin{align}
\Tr\big[\mathcal{E}_{\text{\LON}}\big(D^\dagger(\bm\xi)\big)D(\bm\zeta)\big]
=\pi^M\delta^{2M}\big(\bm\xi-\bm\zeta\bm{L}^\dagger\big)
e^{\bm\zeta(\bm{L}^\dagger\bm L-\bm I_M)\bm\zeta^\dagger}.
\end{align}
Thus the transition function~(\ref{trans-fun}) becomes
\begin{align}
\label{eq:tran-LON}
T_{\text{\LON}}^{(\bm s,\bm t)}(\bm\beta|\bm\alpha)
=\int\frac{\bdm{\zeta}}{\pi^{2M}}\,
e^{-\bm\zeta\bm{\Sigma}\bm\zeta^{\dagger}/2} e^{(\bm\beta-\bm\alpha\bm{L})\bm\zeta^\dagger-\bm\zeta(\bm\beta^\dagger-\bm{L}^\dagger\bm\alpha^\dagger)}
=\frac{2^M}{\pi^M\det\bm\Sigma}
\exp\big[\mathord{-}2(\bm\beta-\bm\alpha\bm L)
\bm\Sigma^{-1}(\bm\beta^\dagger-\bm{L}^\dagger\bm\alpha^\dagger)\big].
\end{align}
\end{widetext}
\noindent
The transition function is well behaved and nonnegative, and has the final (normalized) Gaussian form, if and only if
\begin{align}
\bm\Sigma=\bm{I}_M-\bm{L}^\dagger\bm{L}-\bm s+\bm{L}^\dagger\bm{t}\bm{L}\geq 0,
\end{align}
i.e., $\bm\Sigma$ is positive (semidefinite).  Note that if we choose the same ordering at input and output, i.e., $\bm s=\bm t=s\bm I_M$, then
\begin{align}\label{eq:Sigma11}
\bm\Sigma=(1-s)(\bm{I}_M-\bm{L}^\dagger\bm{L})\ge0,
\end{align}
provided $s\le1$; further choosing $s=1$, we have $\bm\Sigma=0$ and thus \begin{align}\label{eq:Tdelta}
T_{\text{\LON}}^{(\bm{I_M},\bm{I_M})}(\bm\beta|\bm\alpha)=\delta^{2M}(\bm\beta-\bm\alpha\bm L).
\end{align}

To apply our second method for generating an efficient classical simulation of sampling, we should apply the procedure outlined at the end of Sec.~\ref{sec:sufficient}.  Suppose the input state has nonnegative $\TPQD$ $W^{(\bm t)}(\bm\alpha|\bm\rho_\text{in})$ for $\bm t\le\bar{\bm t}$, and the output measurement has nonnegative $\negSPQD$ $W_\Pi^{(-\bm s)}(\bm n|\bm\beta)$ for $\bm s\ge\bar{\bm s}$.  Then the necessary and sufficient condition for our second method to yield an efficient classical simulation of sampling from the output probability distribution is that
\begin{align}\label{eq:barSigma}
\overline{\bm\Sigma}=\bm{I}_M-\bm{L}^\dagger\bm{L}-\bar{\bm s}+\bm{L}^\dagger\bar{\bm{t}}\bm{L}\geq 0.
\end{align}

Two special cases deserve attention.  For a lossless \LON, the transfer matrix $\bm L=\bm U$ is unitary, and the condition~(\ref{eq:barSigma}) becomes $\bar{\bm s}\le\bm{U}^\dagger\bar{\bm{t}}\bm U$.  In the case of \emph{identical\/} measurements on all the output modes, the POVM elements become a product $\Pi_{\bm{n}}=\bigotimes_{k=1}^M \Pi_{n_k}$, where $\{\Pi_{n_k}\}$ is the POVM for the measurement on output mode~$k$, and the $\negSPQD$ of the measurements is also a product,   $W_\Pi^{(-\bm s)}(\bm n|\bm\beta)=\prod_{k=1}^M W_\Pi^{(-s_k)}(n_k|\beta_k)$.  In this situation, the optimal output ordering parameters are the same for all $M$ modes, i.e., $\bar{\bm s}=\bar s\bm I_M$.  Thus, for a lossless \LON\ with identical product measurements, the condition~(\ref{eq:barSigma}) simplifies to $\bar s\bm I_M\le\bm{U}^\dagger\bar{\bm{t}}\bm U$, which is equivalent to $\bar s{\bm I}_M\le\bar{\bm t}$.

In the next two subsections we apply our conditions for classical simulations to two schemes for boson sampling in the presence of errors.  Before turning to that, however, we digress briefly to note that since a \LON\ is a classical process, we can provide a classical simulation for all classical input states, since we can choose $\bm t=\bm s=\bm I_M$, i.e., the $P$~function for the input state and the (always nonnegative) $Q$ function for the measurements; this leads to the $\delta$ transition function of Eq.~(\ref{eq:Tdelta}). This is the motivating case considered at the end of Sec.~\ref{sec:genericsampling}. \ A particular example is provided by inputting coherent states to an \LON\ and performing any measurements at the output.

The flip side of classical input states is classical measurements, such as heterodyne measurements, for which the $P$~functions of the POVM elements are nonnegative, allowing us to choose $\bm s=-\bm I_M$; in this situation, we can choose $\bm t=-\bm I_M$, i.e., the $Q$ function for the input state, and have a nonnegative transition function according to Eq.~(\ref{eq:Sigma11}). Hence, in the case of classical measurements, efficient classical simulation is possible for any input state.

In the symmetric case, where both the input state and the POVM elements have nonnegative Wigner functions, we can choose $\bm t=\bm s=0$, and given Eq.~(\ref{eq:Sigma11}), the transition function is always nonnegative. An example is Gaussian input states to a \LON\ and Gaussian measurements at the output~\cite{BarSan}.

\subsection{Boson sampling with single-photon sources}
\label{sec:BSsinglephotons}

We turn now to investigating the effect of errors in a practical implementation of boson sampling that uses single-photon sources and on-off photodetectors.  Recall that in this model, which is the one proposed originally by Aaronson and Arkhipov~\cite{AA}, $N$ single photons are injected into the first $N$ ports of an $M$-port \LON, with $M\gg N$, and the remaining $N-M$ ports receive the vacuum state.  To avoid having more than one count at an output detector, one generally requires that the number of photons counted at the detectors be $\alt\sqrt M$~\cite{AA,RanSam}.  We consider the following sources of error: impurity of the input photons and mode mismatching of these photons into the \LON, losses and mode mismatching within the \LON, and inefficiency and random counts in the detectors.

It is a considerable practical challenge to generate a single-photon state. We assume that the output of the single-photon sources is a statistical mixture of vacuum and a single photon, $(1-\mu)\ket{0}\!\bra{0}+\mu \ket{1}\!\bra{1}$, $\mu\in[0,1]$.  Note that this state is the output of a beamsplitter with transmissivity $\sqrt{\mu}$ when the beamsplitter is illuminated by a pure single photon.  In addition to the impurity of the input, the input photons are generally not mode-matched to the temporal, frequency, and polarization modes that interfere ideally through the \LON.  The nonoverlapping parts of the input photons are lost to the ideal interference that leads to the probability distribution one wants to sample at the output, so we treat them as a loss and model that loss by virtual beamsplitters with transmissivity $\sqrt{\eta_\ssB}$.  Taking into account both the impurity and the mode mismatching, we have that the state input into the first $N$~ports~is
\begin{align}\label{eq:rhosinglephoton}
\rho=(1-\bar\eta)\ket{0}\!\bra{0}+\bar\eta\ket{1}\!\bra{1},
\end{align}
where $\bar\eta=\mu\eta_\ssB$.  We return to a discussion of mode mismatching, at the input to and throughout the \LON, at the end of this subsection.

By using Eqs.~(\ref{eq:s-PQD-def}) and~(\ref{eq:s-cha-def}) for a single mode, the \tPQD\ of the mixed input state~(\ref{eq:rhosinglephoton}) is given by~\cite{sFock}
\begin{align}\label{eq:tPQDrho}
W^{(t)}(\alpha|\rho)=
\frac{2}{\pi}
\frac{(1-t)(1-t-2\bar\eta)+4\bar\eta|\alpha|^2}{(1-t)^3}e^{-2|\alpha|^2/(1-t)},
\end{align}
which is nonnegative for $t\leq\bar t=1-2\bar\eta=1-2\mu\eta_\ssB$.  As the vacuum state ($\bar\eta=0$) is a classical state whose \tPQD\ is nonnegative for $t\leq1$, the overall \tPQD\ of the input is nonnegative for
\begin{align}
\label{eq:bar-t-sp}
\bm t\leq \bar{\bm{t}}=\bm I_M-2\mu \eta_\ssB\bm{J}_N,
\end{align}
where $\bm J_N$ is the diagonal matrix with 1s in the first $N$ diagonal positions and 0s otherwise.

Losses within the \LON\ are taken into account by the transfer matrix $\bm L$.  For a particular implementation of boson sampling, one should use the measured transfer matrix to analyze the system~\cite{charact}.  A good part of these losses is mode mismatching within the network, about which we say more below.  For our analysis, we adopt a simple model of losses that allows us to investigate how the effect of losses scales with the size of the network.  In particular, we assume that all paths through the \LON{} suffer the same amount of loss and thus describe the network by a transfer matrix $\bm L=\sqrt{\eta_\ssL}\,\bm U$, where $\bm U$ is the unitary transfer matrix for a lossless \LON.  We make this more specific in the following way.  The network consists of $\ell$-port optical elements, each with a uniform transmissivity $\sqrt{\eta_\sszero}$, and has depth $d$.  Thus each input port speaks to $\ell^d$ output ports.  We assume that the network is fully connected, so $M=\ell^d$; hence, each input photon sees a loss $\eta_\ssL=\eta_\sszero^d=\eta_\sszero^{\log_\ell\!M}=M^{\log_\ell\!\eta_0}$.

For the on-off photodetectors, which we assume to be identical, we use a model similar to that devised in Ref.~\cite{Barnett} for detectors with subunity efficiency and dark counts.  We think of the dark-count probability in the model more generally than in the original model, however; it is not just the probability for intrinsic dark counts in the detector, but it also includes any sort of {\it random counts}.  We discuss below how mode-mismatched photons at the input and within the \LON\ can propagate through the \LON\ and contribute random counts at the photodetectors. The POVM elements associated with the on-off outcomes---zero denotes the off state, i.e., no detector click, and 1 denotes a click---are
\begin{align}
\Pi_0(\eta_\ssD,p_\ssD)
&=(1-p_\ssD)\sum_{m=0}^{\infty}(1-\eta_\ssD)^m\ket{m}\bra{m},\\
\begin{split}
\Pi_1(\eta_\ssD,p_\ssD)&=\mathcal{I}-\Pi_0(\eta_\ssD,p_\ssD)\\
&=\sum_{m=0}^{\infty}[1-(1-p_\ssD)(1-\eta_\ssD)^m]\ket{m}\bra{m},
\end{split}
\end{align}
where $\eta_\ssD$, satisfying $0\leq\eta_\ssD\leq 1$, is the detector efficiency, and $1-p_\ssD$ is the probability of no random count.

The sum in $\Pi_0$ is an unnormalized thermal state, so by using the \sPQD\ of a thermal state, we can find the \negsPQD\ of $\Pi_0(\eta_\ssD,p_\ssD)$ to be
\begin{align}
W^{(-s)}_\Pi(0|\beta)
=\frac{1-p_\ssD}{\pi}
\frac{e^{-\eta_\ssD|\beta|^2/[1-\eta_\ssD(1-s)/2]}}{1-\eta_\ssD(1-s)/2};
\end{align}
this is nonnegative provided $s\ge1-2/\eta_\ssD$, which is really no restriction at all.  The \negsPQD\ of $\Pi_1(\eta_\ssD,p_d)$, given by
\begin{align}
W^{(-s)}_\Pi(1|\beta)=\frac{1}{\pi}-W^{(-s)}_\Pi(0|\beta),
\end{align}
is nonnegative provided that $s\ge\bar s=1-2p_\ssD/\eta_\ssD$.  Writing this in terms of the ordering parameters for all the detectors, we have nonnegative output $\SPQD$s~if
\begin{align}
\label{eq:bar-s-sp}
\bm s\geq \bar{\bm s}=\bigg(1-\frac{2p_\ssD}{\eta_\ssD}\bigg)\bm I_M.
\end{align}

Putting Eqs.~(\ref{eq:bar-t-sp}) and~(\ref{eq:bar-s-sp}), plus our description of the \LON, into Eq.~(\ref{eq:barSigma}), we find that the condition for an efficient classical simulation is that
$\bm U\overline{\bm\Sigma}\bm U^\dagger=(2p_\ssD/\eta_\ssD-\eta_\ssL)\bm I_M+\eta_\ssL\bar{\bm t}\ge0$; provided there is even one single-photon input, this reduces to the simple condition
\begin{align}
\label{eq:con-sp}
p_\ssD\ge\eta\equiv\mu\eta_\ssB\eta_\ssL\eta_\ssD=\mu\eta_\ssB\eta_\ssD\eta_\sszero^{\log_\ell\!M},
\end{align}
where $\eta$ characterizes the overall loss in the experiment.  Because of the simplicity of our model for losses within the \LON, the condition~(\ref{eq:con-sp}) does not apply precisely to  \LON{s} with general transfer matrices $\bm L$, but the dependence of $\eta_\ssL$ on $\ell$ and $M$ does indicate how the condition for simulability scales with the size of the~\LON.

A recent study~\cite{Aar-Bro} shows that if a fixed number $K$ of photons are lost, boson sampling remains classically hard, provided $K$ is not too large.  This suggests that in the presence of loss, one can inject more single photons into the \LON\ so that on average, an interesting boson-sampling problem is realized.  The mean number of photodetector counts is $\eta N$; if we require that the number of counts does not exceed $\sqrt M$ much and also require $N\le M$, we have $N=\min(M,\sqrt M/\eta)$.  To get an idea of what is going on, consider an ambitious, but perhaps realistic example in which  $\mu=0.5$, $\eta_\ssB=0.1$, $\eta_\sszero=0.98$, $\ell=2$, and $\eta_\ssD=0.95$.  For $M=10$, we have $\eta_\ssL=0.94$, $\sqrt M/\eta=71$, $N=M=10$, and $N\eta=0.44$; the condition for classical simulability is that $p_\ssD\ge\eta=0.044$.  For $M=100$, we have $\eta_\ssL=0.87$, $\sqrt M/\eta=241$, $N=M=100$, and $N\eta=4.2$; the condition for classical simulability is that $p_\ssD\ge\eta=0.042$.  For $M=1\,600$, we have $\eta_\ssL=0.81$, $\sqrt M/\eta=1\,044=N$, and $N\eta=40$; the condition for classical simulability is that $p_\ssD\ge\eta=0.038$.

An obvious question is why our method needs random counts for classical simulability. The answer is that in the absence of random counts, sampling from the (exact) output probability distribution cannot be efficiently simulated classically; this can be shown using Stockmeyer's approximate counting algorithm.  Even for large losses, it is still possible that all the input photons get counted by the detectors at the output. As any lossy \LON\ can be thought of as part of a larger, lossless \LON, probabilities of these events are proportional to the squared modulus of permanents of complex matrices, which are submatrices of a unitary matrix for the larger \LON. If sampling were classically simulatable, using Stockmeyer's approximate counting algorithm one could approximate one of these probabilities to within a multiplicative error in $\text{BPP}^{\text{NP}}$ [Stockmeyer's algorithm allows for the proportionality factors to be of the order $2^{-\text{poly}(N)}$]; this would lead to the collapse of polynomial hierarchy to the third level because as observed by Aaronson and Arkhipov, multiplicative approximation of these probabilities is \#P hard.  Note, however, that this argument does not imply that boson-sampling experiments with losses and very low random counts are still practically interesting.  One can expect that above some threshold for losses, a classical algorithm can efficiently generate samples from an approximate probability distribution, and in practice, this cannot be distinguished from the outcomes of the experiment.

The importance of random counts prompts us to return to the question of mode mismatching at the input to and within the \LON.  Mode mismatching occurs when temporal, frequency, and polarization properties of photon wave packets do not overlap ideally at the input to the \LON\ and at the optical elements used to implement a specific \LON. The nonoverlapping parts of the photon wave packets are lost to the ideal interference that leads to the probability distribution one wants to sample at the output.  Mode mismatching is thus a loss mechanism and is likely to be the dominant loss mechanism within a large optical network.  Without some intervention, however, nonoverlapping parts of the photon wave packets continue through the \LON\ and are counted within the temporal and spatial windows defined by the photodetectors.  These photocounts are effectively random and contribute to the random-count probability of the detectors (they might be correlated between different output modes, but it is hard to see how this correlation could be used to our advantage); indeed, they are very likely to be the dominant contribution to the random-count probability, as in high-quality detectors, the intrinsic dark-count rate is very low.

In principle, one can use active filters (mode cleaners) to remove nonoverlapping parts of photon wave packets at the input to and output from and perhaps within the \LON\ and, hence, to turn mode mismatching into a genuine loss where the mode-mismatched parts of the photon wave packets do not contribute counts at the photodetectors. To assess how serious this problem is, suppose that mode mismatching is the dominant loss mechanism.  Suppose further that a fraction $f_\ssB$ of the photons lost at the input, numbering $\mu(1-\eta_\ssB)$, continue into the \LON\ and on to the photodetectors and that a fraction $f_\ssL$ of the photons lost within the \LON, numbering $\mu\eta_\ssB(1-\eta_\ssL)$, continue to the detectors.  Assuming these mode-mismatched photons are counted with efficiency $\eta_\ssD$, they contribute random-count probability
\begin{align}\label{eq:pDeff}
p_\ssD
=\frac{\eta_\ssD\mu N}{M}
\Big[f_\ssL(1-\eta_\ssL)\eta_\ssB+f_\ssB(1-\eta_\ssB)\Bigl].
\end{align}
With the same assumptions and same values for loss parameters as above, we now also assume that $f_\ssB=0.1$, on the grounds that the input loss $\eta_\ssB=0.1$ already reflects a major attempt to clean up the input wave-packet modes, and $f_\ssL=0.9$, on the grounds that it would be quite difficult to clean up the output photons without introducing additional losses.  With these assumptions, we get
$p_\ssD=0.046$ for $M=10$,
$p_\ssD=0.049$ for $M=100$, and
$p_\ssD=0.034$ for $M=1\,600$.
Comparing these random counts with the corresponding thresholds from the previous page indicates that mode mismatching is indeed a challenge for boson-sampling experiments of interesting size; recall that this assumes that additional single photons are fed into the \LON\ to compensate for losses in order keep the number of detected photons as close to $\sqrt M$ as possible.  The scaling with $M$ is such that if large \LON{s} can be constructed without compromising the loss parameters, the situation gets better as $M$ increases.

It is worth noting that for the range of parameters we have considered, for which $N\simeq M$, the condition for classical simulatability is that the number of mode-mismatched photons counted at the photodetectors, $Mp_\ssD$, exceed the number
of mode-matched photons, $N\eta$.  This is a useful rule of thumb for assessing the simulatability of a boson-sampling experiment.

\subsection{Boson sampling with SPDC sources}
\label{sec:BSSPDC}

A major practical challenge for implementing boson sampling is reliable single-photon sources.  In most quantum-optics experiments, spontaneous parametric down-conversion (SPDC) is used as a probabilistic source for preparing single photons~\cite{BRO13,SPR13,TIL13,CRE13}.  If the two-mode squeezed vacuum state generated by a SPDC source has weak squeezing, photon counting on the heralding mode prepares vacuum or a single photon in the signal mode, which can then be used as one of the inputs to the $M$ input ports of a boson-sampling \LON.  This scheme can be viewed as sampling from the output photon-counting probability distribution of a larger \LON\ with $2M$ modes; the larger \LON\ consists of the identity process acting on the heralding modes and the original \LON\ acting on the signal modes.  This scenario implements {\it randomized\/} boson sampling, in which when $N$ photons are randomly detected in the heralding modes, $N$ single photons are injected into the corresponding ports of the original~\LON. (With the loss parameters we consider here, in boson sampling with single-photon sources, the single photons are also randomly injected into a LON, but one does not know to which input ports.) In the absence of any losses or inefficiencies, the average number of photons input to the signal-mode \LON\ is $N=M\sinh^2\!r$, where $r$ is the squeezing parameter, assumed to be positive without loss of generality; to achieve $N=\sqrt M\ll M$, one chooses $\sinh^2\!r=1/\sqrt M$~\cite{RanSam}.

We consider the following sources of error: mode mismatching of the signal modes into the smaller, signal-mode \LON, described by virtual beamsplitters with transmissivity $\sqrt{\eta_\ssB}$; losses in the signal-mode \LON, described by the transfer matrix $\bm L$, but no losses for the heralding modes, so that the overall transfer matrix is $\bm I_M\oplus\bm L$; and for all modes, the model for random counts and inefficiency that we introduced previously for on-off detectors.  Since the input squeezed vacuum states are Gaussian and the \LON\ is a Gaussian process, we can efficiently find the \TPQD\ of the output state and use our first condition to check whether efficient classical simulation of the sampling problem is possible.  To simplify our analysis and to compare directly with our results for single-photon inputs, we specialize to the simple model for loss in the signal-mode \LON\ in which $\bm L=\sqrt{\eta_\ssL}\,\bm U$.  Given this assumption, all the signal modes suffer the same loss in the \LON, so we can refer the \LON\ losses to the input, combine them with the mode mismatching of the signal modes, and thus describe both by virtual beamsplitters with transmissivity $\sqrt{\eta_\ssBL}=\sqrt{\eta_\ssB \eta_\ssL}$, which act on each signal mode before it enters the signal-mode~\LON.

The upshot is that the larger \LON\ is fed by $M$ copies of the two-mode state
\begin{align}
\label{eq:2svloss}
\bm\rho_{\textrm{hs}}'
=\Tr_0[\bm\rho_{\textrm{hs}0}]
=\Tr_0 \big[ \mathcal{U}_{s0}(\eta_\ssBL) \bm\rho_{\textrm{hs}}\otimes\ket{0}\!\bra{0} \mathcal{U}_{s0}^{\dagger}(\eta_\ssBL)\big].
\end{align}
Here $\bm\rho_{\textrm{hs}}$ is the two-mode squeezed vacuum state generated by a SPDC source, $\mathcal{U}_{s0}(\eta_\ssBL)$ is the unitary operator for a beamsplitter with transmissivity $\sqrt{\eta_\ssBL}$ that acts on the signal mode of the SPDC and a vacuum input, and the trace is taken over the mode reflected from the beamsplitter.  With the \LON\ losses referred to the input, the larger \LON\ is now described by the unitary transfer matrix $\bm{I}_M \oplus \bm{U}$, which corresponds to a $\delta$-function transfer function that does not alter the negativity of the input \TPQD.

The state $\bm\rho_{\textrm{hs}}'$ is a Gaussian state.  The Wigner function $(t=0)$ of any Gaussian state is a Gaussian function, but if the state is nonclassical, there exists $\bar t\in(0,1]$ such that for $t\le\bar t$, the $\tPQD$ is a Gaussian, for $t=\bar t$, the $\tPQD$ has $\delta$-function singularities, and for $t>\bar{t}$, the \tPQD\ is more singular than a $\delta$ function.  In order to find $\bar t$ for $\bm\rho_{\textrm{hs}}'$, we need to use the covariance matrix of the Gaussian \tPQD\ and find the value of $t$ at which the covariance matrix transitions from positive to negative, i.e., the smallest eigenvalue goes to zero.

The covariance matrix of the Wigner function of $\bm\rho_{\textrm{hs}0}$ in Eq.~(\ref{eq:2svloss}) is given by
 \begin{align}
 \bm\sigma_{\textrm{hs}0}=\big(\bm{I}_2\oplus \bm{B}_{s0}(\eta_\ssBL)\big) \big(\bm\sigma_{\textrm{hs}} \oplus\bm{I}_2\big)\big(\bm{I}_2\oplus \bm{B}^{T}_{s0}(\eta_\ssBL)\big),
 \end{align}
 where
\begin{align}
\bm\sigma_{\textrm{hs}}=
\begin{pmatrix}
\cosh 2r \bm{I}_2 & \sinh 2r \bm{Z}_2 \\[2pt]
\sinh 2r \bm{Z}_2 & \cosh 2r \bm{I}_2 \\
\end{pmatrix}
\end{align}
is the covariance matrix of the two-mode squeezed vacuum state with squeezing parameter $r$, with $\bm{Z}_2=\text{diag}(1,-1)$ being the Pauli $z$ matrix,  and
\begin{align}
\bm{B}_{s0}(\eta_\ssBL)=
\begin{pmatrix}
\sqrt{\eta_\ssBL} \bm{I}_2& -\sqrt{1-\eta_\ssBL} \bm{I}_2 \\[3pt]
\sqrt{1-\eta_\ssBL} \bm{I}_2 & \sqrt{\eta_\ssBL} \bm{I}_2
\end{pmatrix}
\end{align}
\begin{widetext}
\noindent
is the symplectic transformation of a beamsplitter with transmissivity $\sqrt{\eta_\ssBL}$~\cite{Adesso-Illuminati}.  The $4\times 4$ top-left submatrix of $\bm\sigma_{\textrm{hs}0}$ is then the covariance matrix of the Wigner function of $\bm\rho_{\textrm{hs}}'$,
\begin{align}
\label{Sighsp}
\bm\sigma_{\textrm{hs}}'=
\begin{pmatrix}
\cosh 2r \bm{I}_2 & \sqrt{\eta_\ssBL} \sinh 2r \bm{Z}_2 \\[3pt]
\sqrt{\eta_\ssBL} \sinh 2r \bm{Z}_2\; & [1+\eta_\ssBL(\cosh 2r-1)]\bm{I}_2
\end{pmatrix}.
\end{align}

The covariance matrix of the \tPQD\ is given by $\bm\sigma_{\textrm{hs}}'-t\bm{I}_4$; what we need to know is when, as $t$ increases from zero, the smallest eigenvalue of this $4\times4$ matrix goes to zero.  Interchanging rows and columns of $\bm\sigma_{\textrm{hs}}'-t\bm{I}_4$ separates it into the direct sum of two $2\times2$ matrices,
\begin{align}
\begin{split}
&\begin{pmatrix}
\cosh 2r-t& \pm \sqrt{\eta_\ssBL} \sinh 2r \\[3pt]
\pm \sqrt{\eta_\ssBL} \sinh 2r\; & 1-t+\eta_\ssBL(\cosh 2r-1)
\end{pmatrix}\\[3pt]
&\qquad\qquad\qquad
=\big[1-t+(1+\eta_\ssBL)\sinh^2\!r\big]\bm I_2
\pm(1-\eta_\ssBL)\sinh^2\!r\,\bm Z_2\pm2\sqrt{\eta_\ssBL}\sinh r\cosh r\,\bm X_2,
\end{split}
\end{align}
which have the same eigenvalues ($\bm X_2$ is the Pauli $x$ matrix).  The smaller eigenvalue goes to zero when
\begin{align}
\label{eq:bar-t-spdc}
t=\bar t=1+(1+\eta_\ssB\eta_\ssL)\sinh^2\!r
-\sinh r\sqrt{(1+\eta_\ssB\eta_\ssL)^2\sinh^2\!r+4\eta_\ssB\eta_\ssL},
\end{align}
where we have restored $\eta_\ssBL=\eta_\ssB \eta_\ssL$.  For the \TPQD\ of the overall input, we have $\bar{\bm{t}}=\bar{t}\bm{I}_{2M}$.

The analysis of on-off photodetection in Sec.~\ref{sec:BSsinglephotons} shows that nonnegativity of the measurement \SPQD{s} requires $s\ge\bar s=1-2p_\ssD/\eta_\ssD$, and our first method of simulation requires that $s=t\le\bar t$, so the condition for efficient classical simulation of the SPDC scheme is that $\bar s\le\bar t$, which gives
\begin{align}
p_\ssD\ge
-\frac12\eta_\ssD(1+\eta_\ssB\eta_\ssL)\sinh^2\!r
+\frac12\eta_\ssD\sinh r\sqrt{(1+\eta_\ssB\eta_\ssL)^2\sinh^2\!r+4\eta_\ssB\eta_\ssL}.
\end{align}

\end{widetext}

The mean number of photons input to the signal modes is $N=M\sinh^2\!r$, meaning that $\sinh^2\!r$ in the above expressions is a surrogate for $N/M$.  The average number of counts at the photodetectors is $\eta' N=\eta' M\sinh^2\!r$, where $\eta'=\eta_\ssD \eta_\ssL\eta_\ssB$ gives the total loss through the system.  As in our analysis of single-photon boson sampling, we choose $\eta' N=\sqrt M$ provided that $N\le M$; i.e., we choose $N=\min(M,\sqrt M/\eta')$, which is equivalent to $\sinh^2\!r=\min(1,1/\sqrt M \eta')$.  Again we consider experiments in which $\eta_\ssB=0.1$, $\eta_\sszero=0.98$, $\ell=2$, and $\eta_\ssD=0.95$.  For $M=10$, we have $\eta_\ssL=0.94$, $\sqrt M/\eta'=36$, $N=M=10$, $N\eta'=0.89$, and $\sinh^2\!r=1$; the threshold for classical simulability is $p_\ssD\ge 0.076$.  For $M=100$, we have $\eta_\ssL=0.87$, $\sqrt M/\eta'=120$, $N=M=100$, $N\eta'=8.3$, and $\sinh^2\!r=1$; the threshold for classical simulability is $p_\ssD\ge 0.071$.  For $M=1\,600$, we have $\eta_\ssL=0.81$, $\sqrt M/\eta'=522=N$, $N\eta'=40$, and $\sinh^2\!r=0.33$; the threshold for classical simulability is $p_\ssD\ge 0.060$.  It is notable that these thresholds are close to twice those we found under comparable conditions for single-photon boson sampling; the single-photon thresholds would be higher if the single-photon sources produced photons with no impurity ($\mu=1$).

As in single-photon boson sampling, SPDC boson sampling suffers from the problem of mode-mismatched photons becoming random counts in the photodetectors.  The same analysis as for single-photon boson sampling yields random-count probability~(\ref{eq:pDeff}) with $\mu=1$.  Again assuming $f_B=0.1$ and $f_L=0.9$, with all the other parameters the same as above, the random-count probability is
$p_\ssD=0.091$ for $M=10$,
$p_\ssD=0.096$ for $M=100$, and
$p_\ssD=0.033$ for $M=1\,600$.  This indicates that mode mismatching is a challenge for SPDC boson-sampling experiments of interesting size.  Just as for single-photon sources, this conclusion assumes that additional photons are input to compensate for losses, but again the scaling is favorable provided one can keep losses and mode mismatching under control as system size increases.

\section{Conclusion}
\label{sec:conclusion}

In this paper we established sufficient conditions for efficient classical simulation of general quantum-optical experiments that involve a quantum state that is subjected to an $M$-mode quantum process and measurement at the output of the process.  These conditions support the notion that negativity is a quantum resource by showing that efficient classical simulation of sampling from the output probability distribution is possible when there are (i)~nonnegative output-state and output-measurement quasiprobability distributions or (ii)~nonnegative input-state and output-measurement quasiprobability distributions and a nonnegative transition function associated with the quantum process.

We applied our conditions for classical simulability to two implementations of the boson-sampling problem.  We considered simple models of errors and imperfections to assess the effects of mode mismatching, loss in the \LON, and inefficiency and random counts of on-off photodetectors.  We found that these errors have a significant impact and obtained random-count thresholds beyond which efficient classical simulation is possible. For any actual implementation of boson sampling, however, one should go beyond the simple examples given here and use our methods to model all the imperfections, noise, and errors, particularly, formulating and analyzing a detailed model of losses and mode mismatching within the particular \LON, in order to determine when it is possible to do classical simulations using our methods.  In the case of mode mismatching, nonoverlapping parts of photon wave packets that proceed to and are counted at the detectors are likely to be the major contribution to the random-count probability; hence, it is particularly important to assess the need for and effectiveness of active mode cleaning (so-called quantum filters) to mitigate this effect.

We caution that we do not warrant that there is no other method of efficient classical simulation when our conditions are not satisfied.  Indeed, we have only considered the problem of sampling from the exact output probability distribution of measurement outcomes.  A more general problem is approximate sampling, i.e., sampling from a close approximation to the exact probability distribution, in which case the question is whether sampling from the approximate distribution can be efficiently simulated classically.  We have shown that in the presence of losses in boson-sampling experiments and with zero or very low random counts, the exact sampling problem cannot be simulated using our methods.  Yet, under the same conditions, one might be able to simulate approximate sampling.  A possible approach might be to simulate sampling from a nonnegative distribution that approximates a slightly negative quasidistribution, perhaps using techniques like those recently introduced for discrete-variable systems~\cite{Pashayan15}.  We leave this as a subject for future research.

Several lessons might be drawn from our work in this paper.  First, in any protocol that uses probabilistic state preparation, the state preparation should be included when one searches for efficient classical simulations.  If classical simulation is possible for sampling from the whole distribution, then it is also possible for sampling from a subdistribution that is chosen by postselection.  This is the approach we used in our analysis of SPDC boson sampling, where we included the heralding modes explicitly in the search for a classical simulation.  Second, our random-count thresholds are hard boundaries.  These hard boundaries might be moved closer to the ideal problem by considering approximate sampling, as discussed above, but the point here is that such hard boundaries might not be found by considering perturbations about an ideal protocol.  This might be a general property of analogue quantum protocols like boson sampling.  A third lesson is that it is generally easier to devise analogue quantum protocols than it is to show that the protocol does not have an efficient classical simulation.  Confronted with a new analogue quantum protocol, the responsibility of theorists and experimenters alike is to put on the classical thinking cap and to focus on whether classical simulations are possible in the presence of noise; this is essential for designing experiments that are meaningful implementations of the quantum protocol.

Our methods for classical simulation are based on the wave aspects of boson-sampling experiments, as opposed to the particle aspects.  One classical analogue of boson sampling replaces the identical input bosons with classical distinguishable particles undergoing probabilistic transitions within a network; in this situation, output probabilities are given by permanents of matrices with nonnegative elements~\cite{AA}.  In contrast, in our methods, we deal with waves undergoing interference within a \LON\ and try to mimic quantum mechanics by using quasiprobability distributions to translate from particle inputs and particle measurements to the complex amplitudes of interfering waves.  This is a natural way to try to simulate an analogue quantum protocol like boson sampling.  We close by noting that the mode-mismatched photons that make their way to the detectors, which we identify as the chief challenge for boson sampling, are effectively the distinguishable photons of a particle description.

\acknowledgments
We thank F.~Shahandeh, A.~Lund, W.~Bowen, M.~Bremner, A.~Fedrizzi, M.~Almeida, and J.~Loredo for informative discussions. This research was supported in part by the Australian Research Council Centre of Excellence for Quantum Computation and Communication Technology (Project No.~CE110001027) and by U.S. National Science Foundation Grants No.~PHY-1314763 and No.~PHY-1521016.

\end{document}